\documentclass[prb,aps,twocolumn]{revtex4}
\usepackage{amsmath}
\usepackage{amssymb}
\usepackage{bm}
\usepackage{epsf}
\begin{document}

\title{Spin injection and detection by resonant
tunneling structure}
\author{M.M.~Glazov, S.A.~Tarasenko, P.S.~Alekseev,
M.A.~Odnoblyudov, V.M.~Chistyakov, and I.N.~Yassievich}

\affiliation{A.F.~Ioffe Physico-Technical Institute, RAS, 194021
St.~Petersburg, Russia}

\begin{abstract}
{A theory of spin-dependent electron transmission through resonant
tunneling diode (RTD) grown of non-centrosymmetrical semiconductor
compounds has been presented. It has been shown that RTD can be
employed for injection and detection of spin-polarized carriers:
(i) electric current flow in the interface plane leads to spin
polarization of the transmitted carriers, (ii) transmission of the
spin-polarized carriers through the RTD is accompanied by
generation of an in-plane electric current. The microscopic origin
of the effects is the spin-orbit coupling-induced splitting of the
resonant level.}
\end{abstract}

\maketitle

\section{Introduction}

Spin-dependent phenomena and particularly transport of
spin-polarized carriers in semiconductor heterostructures attract
a great attention. One of the key problems of spintronics is a
development of efficient methods of injection and detection of
spin-polarized carriers. Among various techniques ranging from
optical orientation to spin injection from magnetic materials,
development of non-magnetic semiconductor injectors and detectors
is of the special interest. Spin-orbit interaction underlying such
devices couples spin states and space motion of conduction
electrons and makes conversion of electric current into spin
orientation and vice versa possible.

Recently it was demonstatred that transparency of semiconductor
barriers may depend on the spin orientation of carriers. Two
microscopic mechanisms, Rashba spin-orbit coupling induced by the
barrier asymmetry~[1-3] and the $k^3$ Dresselhaus spin splitting
in non-centrosymmetrical materials~[4-6], were shown to be
responsible for the effect of spin-dependent tunneling. Devices
based on spin-dependent tunneling were suggested to be utilized as
components of the spin field effect transistor~[7].

In this report we present a theory of resonant spin-depen\-dent
tunneling through the RTD structure grown of non-centrosymmetrical
semiconductors. We show that the RTD can be employed for injection
and detection of spin-polarized carriers. The effects originate
from the spin-orbit coupling-\-induced splitting of the resonant
level.

\section {Resonant spin-dependent tunneling}

We consider the transmission of electrons with the initial
wavevector $\bm{k}=(\bm{k}_{\parallel},k_z)$ through a symmetrical
double barrier structure grown along $z \parallel [001]$ direction
(see Fig.1). Here $\bm{k}_{\parallel}$ is the wavevector in the
plane of the interfaces, and $k_z$ is  wavevector component normal
to the barrier and pointing in the direction of tunnelling. The
effective electron Hamiltonian of non-centrosymmetrical
semiconductors contains spin-dependent $k^3$ Dresselhaus term
that, for the case of the RTD structure, can be simplified to
\begin{equation}
\mathcal H_{D} = \gamma (\hat{\sigma}_x k_x - \hat{\sigma}_y k_y)
\frac{\partial^2}{\partial z^2}  \:,
\end{equation}
where $\gamma$ is a constant depending on the compound,
$\hat{\sigma}_{\alpha}$ are the Pauli matrices, and the coordinate
axes $x,y,z$ are assumed to be parallel to the cubic
crystallographic axes $[100]$, $[010]$, $[001]$, respectively.
\begin{figure}[t]
\vspace{10mm} \leavevmode  \epsfxsize=3in
\centering{\epsfbox{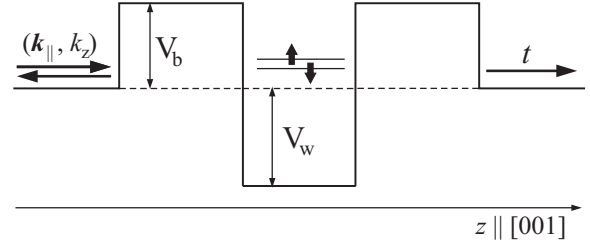}}
\vspace{-2mm}
\caption{Electron transmission through RTD structure.}
\end{figure}
The electron spins $\bm{s}_\pm$ corresponding to the spin
eigen-states "$\pm$" of the Dresselahaus term~(1) depend on the
electron wavevector and are given by
\begin{equation}
\bm{s}_{\pm} = 1/2 \, (\mp \cos \varphi \,, \: \pm \sin \varphi
\,, \: 0) \:,
\end{equation}
where $\varphi$ is the polar angle of the wavevector $\bm{k}$ in
the $xy$ plane, $\bm{k}_\parallel = (k_{\parallel} \cos \varphi
\,, \: k_{\parallel} \sin \varphi)$.

Solution of the Schr\"odinger equation with the term $\mathcal
H_{D}$ included allows one to derive the transmission, $t_\pm$,
and reflection, $r_\pm$, coefficients for the electrons of the
spin states "$\pm$". Fig.2 presents the dependencies of the RTD
transparency, $|t_\pm|^2$, on the incident electron energy along
growth direction, $E_z=\hbar^2 k^2_z / 2 m^*$, at certain in-plane
wave\-vector $k_\parallel$. The spin splitting of the resonant
peak is clearly seen.
\begin{figure}[h]
\leavevmode \epsfxsize=3in \centering{\epsfbox{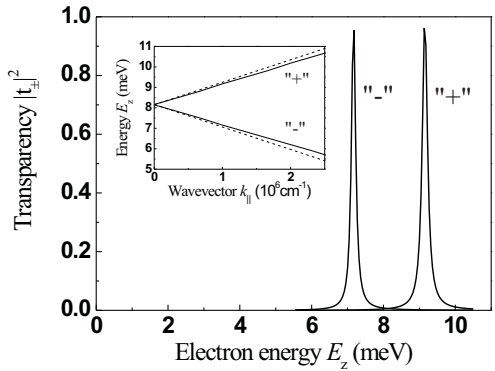}}
\vspace{3mm}
\caption{The RTD transparency, $|t_\pm|^2$, as a function of $E_z$
at fixed $k_\parallel = 10^6$~cm$^{-1}$. The insert shows the
dependence of the spin splitting of the resonant peak on
$k_\parallel$ calculated numerically (solid curves) and following
Eq.(4) (dashed lines). The used parameters, $\gamma =
76$~eV$\cdot$\AA$^{3}$, $m^*=0.053m_0$, $V_{{\rm b}}=230$~meV,
$V_{{\rm w}}=200$~meV, $a=29$\AA, and $b=60$\AA, correspond to
Al$_{x}$Ga$_{1-x}$Sb, $x=0.15/0.3/0/0.3/0.15$, RTD structure.}
\end{figure}

In the case of thick barriers the structure transparency
demonstrates a sharp peaks and hence can be approximated by
$\delta$-functions
\begin{equation}
\left|t_\pm(E_z, k_{\|})\right|^2 \, \approx \, \pi \Gamma_\pm({
k_\|})\, \delta\left[E_z-E_{\pm}( k_{\|})\right],
\end{equation}
where transmission efficiency, $\Gamma_\pm({ k_\|})$, and the
resonance energy, $E_{\pm}( k_{\|})$, can be expanded as
\begin{equation}
E_{\pm}({ k}_\parallel) \approx E_0 \pm \alpha k_\parallel \:,
\quad \Gamma_{\pm}({ k}_\parallel) \approx ( 1 \pm \beta
k_\parallel)\Gamma_0 \:.
\end{equation}
Here $E_0$ and $\Gamma_0$ are the position and the width of the
level when the spin-orbit interaction is neglected, the constants
$\alpha$ and $\beta$ are given by
\[
\alpha= \frac{2\gamma m^*}{\hbar^2} \frac{V_{{\rm w}} + E_0}{1 +
2/\kappa a}\:, \quad \beta = 2\frac{b m^*}{\hbar^2\kappa} (\alpha
+ \gamma \kappa^2) \:,
\]
$a$ and $b$ are the well width and the barrier thickness, $\kappa=
\sqrt{2m^*(V_{{\rm b}}-E_0)}/\hbar$ is the electron wavevector
under the barrier, $V_{{\rm w}}$ and $V_{{\rm b}}$ are the well
depth and the barrier height, respectively, and $m^*$ is the
effective mass.

\section {Spin injection}

Spin splitting of the resonant peak at non-zero $\bm
k_{\parallel}$ can be employed for injection of spin-polarized
carriers. We consider two parts of bulk semiconductor separated by
a RTD structure. In equilibrium the momentum distribution of
incident electrons is isotropic and therefore the average spin of
transmitted carriers vanishes. This isotropy can be broken by
application, for example, of an in-plane electric field, $\bm F$.
The carriers tunnel now with non-zero average wavevector in the
plane of interfaces that leads to the spin polarization of
transmitted electrons.

The degree of spin polarization of electrons transmitted through
the RTD has the form
\begin{equation}
P_s = \frac{v_d \,m^*}{\hbar} \left( \alpha / \varepsilon - \beta
\right)\:,
\end{equation}
where $v_{d}=(e \tau_p / m^*) F$ is the drift velocity of the
incident particles, $e$ is the electron charge, $\tau_p$ is the
momentum scattering time, and $\varepsilon$ is an energy equal to
$E_F-E_0$ for 3D Fermi and $k_B T$ for 3D Boltzmann gas,
respectively.
\begin{figure}[b]
\leavevmode  \epsfxsize=3.5in
\centering{\epsfbox{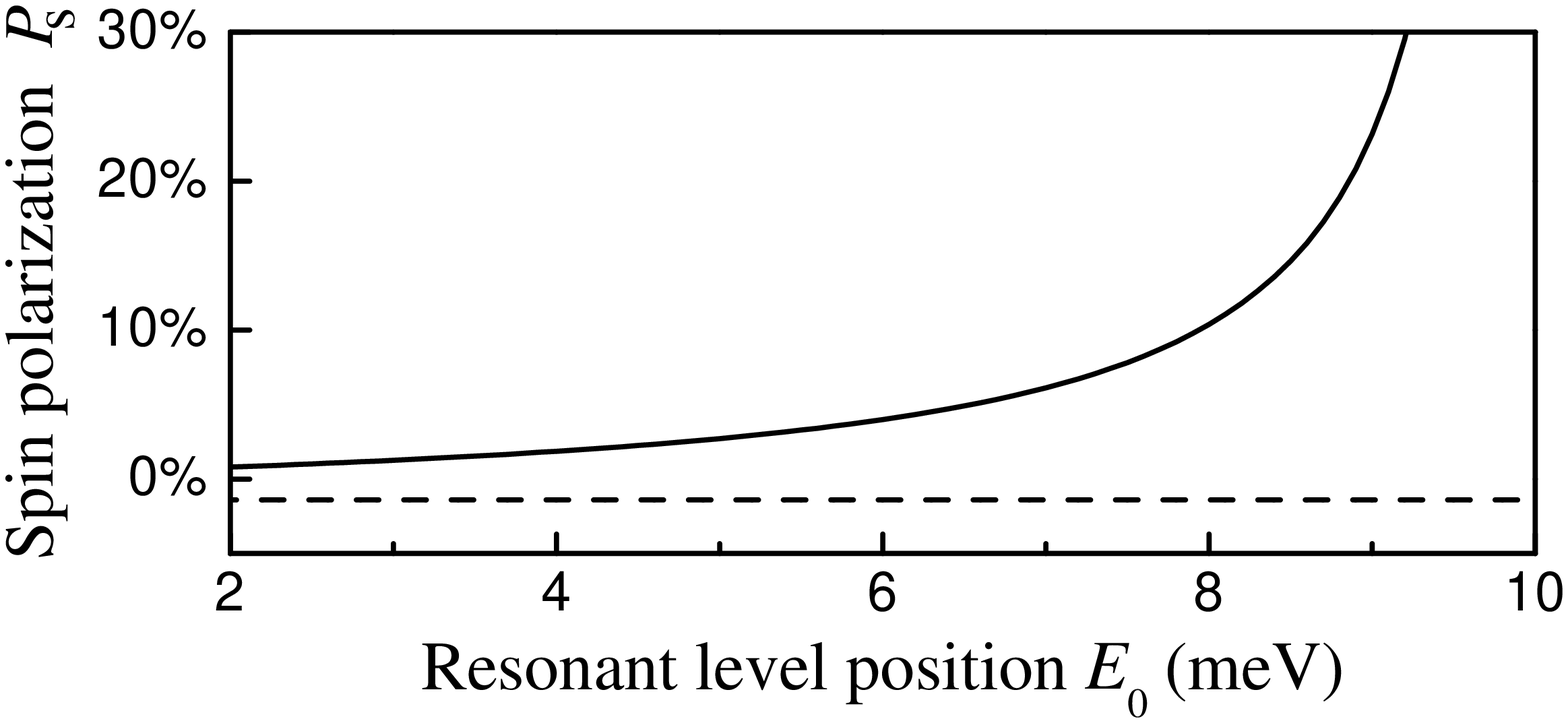}}
\vspace{-3mm}
\caption{The spin polarization, $P_s$, as a function of the
resonant level position, $E_0$, for Fermi gas with $E_F=10$~meV
(solid curve) and Boltzmann gas at $T=300$~K (dashed line). The
parameters of the double-barrier structure are presented in
caption to Fig.2, and $v_d=5\cdot10^6$~cm/s.}
\end{figure}

The dependence of the spin polarization, $P_s$, on the resonant
level position, $E_0$, is presented in Fig.3 for generate and
non-generate 3D electron gas. The change of the sign of $P_s$ with
the temperature increasing is related to interplay between the
spin splitting, $\alpha k_{\parallel}$, and modification of the
widths of the resonant spin sublevels, $\beta k_{\parallel}$.

\section {Tunneling spin-galvanic effect}

Generation of an electric current by spin-polarized carriers
represents the effect inverse to spin injection. Now we assume the
electron gas on the left side of the double-barrier structure to
be spin-polarized. Electrons with various wave\-vectors tunnel
through the barrier. However due to the splitting of the resonant
level, the RTD transparency for the spin-polarized carriers with
the certain in-plane wavevector $\bm{k}_{\parallel}$ is larger
than the transparency for the particles with the opposite in-plane
wavevector, $-\bm{k}_{\parallel}$. This asymmetry results in the
in-plane flow of the transmitted electrons near the barrier, i.e.
in the interface electric current.

The interface current generated by transmission of spin-polarized
electrons through the RTD structure is given by
\begin{equation}
j_{\parallel} = \frac{e\tau_p \, m^*}{\pi \hbar^4}
\frac{\Gamma_0}{\langle 1/E \rangle} f (E_0) \left(\alpha - \beta
\varepsilon\right)  P_s \:,
\end{equation}
where $\langle 1/E \rangle$ is the average value of the reciprocal
kinetic energy of the carriers, equal to $3/E_F$ for degenerate
and $2/ k_B T$ for non-degenerate 3D electron gas, and $f(E)$ is
the carrier distribution function.

Fig.4 presents the dependence of the interface current,
$j_{\parallel}$, on the energy position of the resonant level,
$E_0$, for degenerate and non-degenerate electron gas. The
direction of the current is determined by the spin orientation of
the incident electrons with respect to the crystal axes. In
particular, the current $\bm{j}_{\parallel}$ is parallel (or
antiparallel) to the electron spin, $\bm{s}$, if $\bm{s}$ is
directed along axes $[100]$ or $[010]$; and $\bm{j}_{\parallel}$
is perpendicular to $\bm{s}$, if it is directed along $[1
\bar{1}0]$ or $[1 10]$.
\begin{figure}[t]
\vspace{-5mm} \leavevmode  \epsfxsize=3.5in
\centering{\epsfbox{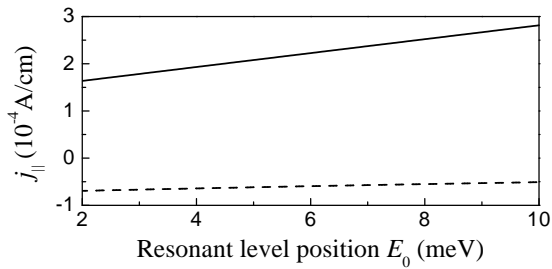}}
\vspace{-20mm}
\caption{The interface current, $j_\parallel$, as a function of
the resonant level position, $E_0$, for Fermi gas with
$E_F=10$~meV (solid curve) and Boltzmann gas at $T=300$~K (dashed
line). The parameters are presented in caption to Fig.2, $\tau_p =
1$~ps, and $P_s=0.1$.}
\end{figure}

In conclusion, the theory of spin-dependent tunneling has been
developed for double-barrier structures based on
non-centrosymmetrical semiconductors. It has been shown that RTD
could be employed for injection and detection of spin-polarized
carriers.

\begin{acknowledgements}
 This work was supported by the RFBR, the INTAS, programs of
the RAS, and Foundation "Dynasty" - ICFPM.
\end{acknowledgements}

\end{document}